\documentclass[prl,twocolumn,showpacs,lengthcheck]{revtex4}

\usepackage{amsmath}
\usepackage{amssymb}
\usepackage{amsthm}
\usepackage{graphicx}

\newcommand{\tr}{\operatorname{tr}}
\newcommand{\uinvnorm}{|\kern-2pt|\kern-2pt|}

\theoremstyle{plain}

\theoremstyle{definition}

\theoremstyle{remark}

\begin{document}
\bibliographystyle{apsrev}

\title{A Renormalisation-Group Algorithm for Eigenvalue Density Functions of Interacting Quantum Systems}

\author{Tobias J.\ Osborne}
\email[]{Tobias.Osborne@rhul.ac.uk} \affiliation{Department of
Mathematics, Royal Holloway University of London, Egham, Surrey TW20
0EX, United Kingdom}

\date{\today}

\begin{abstract}
We present a certifiable algorithm to calculate the eigenvalue
density function --- the number of eigenvalues within an
infinitesimal interval --- for an arbitrary $1$D interacting quantum
spin system. Our method provides an arbitrarily accurate numerical
representation for the \emph{smeared} eigenvalue density function,
which is the convolution of the eigenvalue density function with a
gaussian of prespecified width. In addition, with our algorithm it
is possible to investigate the density of states near the ground
state. This can be used to numerically determine the size of the
ground-state energy gap for the system to within a prespecified
confidence interval. Our method exploits a finitely correlated
state/matrix product state representation of the propagator and
applies equally to disordered and critical interacting $1$D quantum
spin systems. We illustrate our method by calculating an
approximation to the eigenvalue density function for a random
antiferromagnetic Heisenberg model.
\end{abstract}


\maketitle

%
%

The statics and dynamics of interacting quantum many-particle
systems are still relatively poorly understood. Indeed, even
calculating an approximation to such basic quantities as the
ground-state energy appears to be extremely difficult for many
interesting systems. At least one reason for this is that for
arbitrary local quantum systems this problem is complete for the
complexity class {\sf QMA}, which is the quantum analogue of {\sf
NP} \cite{kitaev:2002a, kempe:2004a, oliveira:2005a}. Of course, we
do not really expect that there exist general efficient
computational schemes to study eigenvalues and related thermodynamic
properties. But, it is plausible that for \emph{realistic} quantum
systems there may exist efficient schemes to calculate certain
physical properties like \emph{approximations} to energy gaps and
other thermodynamic properties.

The development of the density matrix renormalisation group (DMRG)
has provided us with what promises to be an efficient way to
calculate physical properties of the ground states of interacting
quantum systems in $1$D (see \cite{schollwoeck:2005a} and references
therein for a detailed description of the DMRG). While the method
was originally developed to obtain approximations to the ground
state of a regular interacting quantum spin lattice system in $1$D,
the DMRG is an extremely flexible method and has been recently
extended to apply to a diverse number of situations, such as the
calculation of short-time dynamics \cite{vidal:2003a, vidal:2003b},
dissipation \cite{verstraete:2004b, zwolak:2004a}, eigenstates with
definite momentum \cite{porras:2005a}, and, recently, higher
dimensions \cite{verstraete:2004a}.

Whether the DMRG and related algorithms actually compute
approximations to the ground state of a quantum system instead of
low-lying excited states is an open question. This problem is
difficult to answer because the DMRG cannot be \emph{certified},
i.e., once the DMRG produces a ground-state approximation there is
no way to prove that this approximation is correct to within some
prespecificed confidence interval. However, this situation is
changing; there have recently been several works which provide
certifiable DMRG-like algorithms to approximate ground states of
interacting spin systems \cite{osborne:2005d, osborne:2005e,
osborne:2006a, osborne:2006b, osborne:2006c, verstraete:2005a,
hastings:2005a}.

One situation where DMRG-related algorithms have been less
successful is in the calculation of the \emph{eigenvalue density
function} $\mu_H(x)$ and the \emph{eigenvalue counting function}
$N_H(x)$, which counts the number of eigenvalues of a hamiltonian
$H$ with value less than $x$ \cite{endnote21}. (The two functions
are connected by $\mu_H(x) = dN_H(x)/dx$.) Perhaps the closest
general method which has been developed along these lines is due to
Porras, Verstraete, and Cirac \cite{porras:2005a}. This method
calculates eigenstates with definite linear momentum for $1$D
quantum spin systems on a ring. However, it is possible for the
method of Porras, Verstrate, and Cirac to miss eigenstates in the
same way that the DMRG can miss the ground state and end up in a
local minima. While this never appears to occur in practice it would
be desirable to have a method which trades this uncertainty against
some other approximation. Additionally, if the method of Porras,
Verstraete, and Cirac is used to calculate $\mu_H(x)$ for $x \gg
O(1)$ above the ground state energy then this would require
exponential resources.

In this Letter we introduce a certifiable method to calculate
systematic approximations to $\mu_H(x)$ for rather general $1$D
quantum spin systems (a generalisation to two and higher dimensions
is available, which uses a slight modification of the technology of
Verstraete and Cirac \cite{verstraete:2004a} and
\cite{richter:1996}). We assume neither translation invariance nor
noncriticality. Our method is an approximation because it calculates
a ``smeared'' version $\widetilde{\mu}_H(x)$ of $\mu_H(x)$, which is
the convolution of $\mu_H(x)$ with a gaussian which has a width
which can be reduced with a complexity that provably scales
polynomially with $n$. Our method doesn't suffer from the
uncertainty of the method of Porras, Verstraete, and Cirac, i.e.,
that maybe some eigenstates are missed. However, the price we pay
for this is that while it is certain that every eigenvalue is
represented in $\widetilde{\mu}_H(x)$ there is some inevitable
uncertainty in the calculated positions of the eigenvalues.

The outline of this Letter is as follows. We begin by introducing
some definitions and we introduce the class of systems we study. We
then describe our numerical renormalisation group method to
calculate $\widetilde{\mu}_H(x)$. We conclude with some numerical
results of our method applied to a random antiferromagnetic
Heisenberg model.

\begin{figure}
\center
\includegraphics{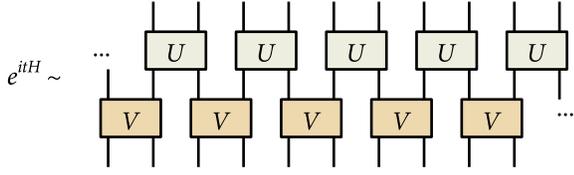}
\caption{The structure of the propagator $e^{itH}$ for a quantum
spin system for times $t$ which are short in comparison to the size
of the system (see \cite{osborne:2005d, eisert:2006b, bravyi:2006a}
for additional discussion).}\label{fig:qca}
\end{figure}

%
%

We will, for the sake of clarity, introduce and describe our method
for a chain of $n$ distinguishable spin-$\frac{1}{2}$ particles.
Thus, the Hilbert space $\mathcal{H}$ for our system is given by
$\mathcal{H} = \bigotimes_{j=0}^{n-1} \mathbb{C}^2$. Consider the
$C^*$-algebra $\mathcal{B}(\mathcal{H})$ which is the Hilbert space
of all (bounded) linear operators $A$ on $\mathcal{H}$ with inner
product $(A, B) = \tr(A^\dag B)$. An orthonormal basis for
$\mathcal{B}(\mathcal{H})$ is given by by $
\sigma^{\boldsymbol{\alpha}} = \sigma^{\alpha_0}\otimes
\sigma^{\alpha_1} \otimes \cdots \sigma^{\alpha_{n-1}}$,
$\alpha_j\in \mathbb{Z}/4\mathbb{Z}$, $0\le j\le n-1$, which we call
the \emph{standard operator basis}, where $
\sigma^{\alpha} = \left[ \left(\begin{smallmatrix} 1 & 0 \\
0 & 1
\end{smallmatrix}\right), \left(\begin{smallmatrix} 0 & 1 \\ 1 & 0
\end{smallmatrix}\right), \left(\begin{smallmatrix} 0 & -i \\ i & 0
\end{smallmatrix}\right), \left(\begin{smallmatrix} 1 & 0 \\ 0 & -1
\end{smallmatrix}\right) \right],
$ is the vector of Pauli sigma matrices. We write the structure
constants ${g^{\alpha\beta}}_{\gamma}$ for the $C^*$-algebra
generated by $\sigma^{\alpha}$: $\sigma^{\alpha}\sigma^{\beta} =
\sum_{\gamma=0}^3{g^{\alpha\beta}}_{\gamma}\sigma^{\gamma}$. The
family $H$ of local hamiltonians we focus on is defined by $H =
\sum_{j=0}^{n-1} h_j$, where $h_j$ is an interaction term which
couples only neighbouring spins $j$ and $j+1$. We assume the
standard energy normalisation whereby $\|h_j\| = O(1)$. The
interaction $h_j$ may vary with position.

%
%

The objective of this Letter is to understand the distribution of
eigenvalues for the operator $H$. To do this we'll study the
\emph{eigenvalue density function} $\mu_{H}(x)$ for $H$ which is
given by
\begin{equation}
\mu_{H}(x)=\frac{1}{2^n}\sum_{j=0}^{2^n-1} \delta(E_j-x),
\end{equation}
where $\delta(x)$ is the Dirac delta function and $E_j$ are the
eigenvalues of $H$. The \emph{eigenvalue counting function}
$N_{H}(x)$ of $H$ is defined to be equal to the number of
eigenvalues of $H$ which are less than or equal to $x$.

The eigenvalue density function $\mu_{H}(x)$ for an operator $H$ has
a delta function spike at the position of each eigenvalue of $H$.
Notice that we have normalised the eigenvalue density function
$\mu_{H}(x)$ to have area $1$, i.e.\ $\int_{-\infty}^{\infty} dx\,
\mu_{H}(x) = 1$. We have done this principally so that we can
compare the eigenvalue densities for operators on different Hilbert
spaces. The eigenvalue counting function $N_{H}(x)$ can be expressed
in terms of the eigenvalue density function $\mu_{H}(x)$ as
$N_{H}(x) = 2^n\int_{-\infty}^x d\omega\,\mu_{H}(\omega)$.

We obtain the eigenvalue density function $\mu_{H}(x)$ for an
operator ${H}$ via the following procedure. Write ${H}$ in its
eigenbasis, ${H} = \sum_{j=0}^{2^n-1} E_j |E_j\rangle \langle E_j|$,
and consider the \emph{propagator} $U(t) = e^{it {H}}= \sum_{j =
0}^{2^n-1} e^{iE_jt} |E_j\rangle \langle E_j|$. Taking the fourier
transform of the scaled propagator $\frac{1}{2^n}U(t)$ yields
\begin{equation}
\frac{1}{2^n}\widehat{U}(\omega) = \frac{1}{2^n}\mathfrak{F}[U(t)]
= \frac{2\pi}{2^n}\sum_{j=0}^{2^n-1} \delta(E_j-\omega)
|E_j\rangle \langle E_j|,
\end{equation}
where the fourier transform pair $(\mathfrak{F}[\cdot],
\mathfrak{F}^{-1}[\cdot])$ is defined to be $F(\omega) =
\mathfrak{F}[f(t)] = \int_{-\infty}^{\infty} dt\, f(t)e^{-i \omega
t}$ and $f(t) = \mathfrak{F}^{-1}[F(\omega)] =
\frac{1}{2\pi}\int_{-\infty}^{\infty} dt\, F(\omega)e^{i \omega t}$.
If we take the trace of $\frac{1}{2^n}\widehat{U}(\omega)$ we find
$2\pi\mu_{H}(\omega) = \frac{1}{2^n}\tr(\widehat{U}(\omega)) =
\frac{2\pi}{2^n}\sum_{j=0}^{2^n-1} \delta(E_j-\omega)$.

The calculations in the previous paragraph show that if we know
$\tr(U(t))$ for arbitrary times then we have enough information to
extract $\mu_{H}(x)$. Now we show that if we only know an
approximation $V(t)$ to $U(t)$ valid for some time $|t| \le T$ then
we can still extract an approximation $\nu_{H}(x)$ to $\mu_{H}(x)$
which can be systematically improved as $T$ is increased. The idea
is to introduce a scalar-valued windowing function $\chi_T(t)$ which
cuts off the propagators $U(t)$ and $V(t)$ outside $|t| \le T$ so
that $\chi_T(t)V(t) \sim \chi_T(t)U(t)$ for all $t$. One convenient
choice for $\chi_T(t)$, which we use in the sequel, is the gaussian:
\begin{equation}\label{eq:chigauss}
\chi_T(t) = \frac{e^{-\frac{t^2}{2T^2}}}{\sqrt{2\pi}T}.
\end{equation}

If we now study $\frac{1}{2^n}\chi_T(t)\tr(U(t))$, rather than
$\frac{1}{2^n}\tr(U(t))$, then a fourier transform and an
application of the convolution theorem yields
\begin{equation}\label{eq:smoothedft}
\frac{1}{2^n}\mathfrak{F}[\chi_T(t)\tr(U(t))] =
\frac{2\pi}{2^n}\sum_{j=0}^{2^n-1} \widehat{\chi}_T(E_j-\omega),
\end{equation}
where $\widehat{\chi}_T(\omega)$ is the fourier transform of the
windowing/characteristic function. It is straightforward to
identify Eq.~(\ref{eq:smoothedft}) as a convolution
\begin{equation}
\frac{1}{2^n}\mathfrak{F}[\chi_T(t)\tr(U(t))] =
2\pi(\widehat{\chi}_T\star\mu_{H})(\omega).
\end{equation}
In this way we identify the fourier transform of
$\frac{1}{2^n}\chi_T(t)\tr(U(t))$ with a \emph{smearing} of the
eigenvalue density function $\mu_{H}(\omega)$ with a smearing
function $\widehat{\chi}_T(\omega)$. When $\chi_T(t)$ is chosen to
be a gaussian, as in Eq.~(\ref{eq:chigauss}), it is clear that
increasing the time window $T$ reduces the width of
$\widehat{\chi}_T(\omega)$. Thus, in the limit $T\rightarrow\infty$
we smoothly (but not uniformly!) recover the eigenvalue distribution
function: $\mu_{H}(\omega) = \lim_{T\rightarrow\infty}
(\widehat{\chi}_T\star\mu_{H})(\omega)$.

It is immediate that if we now have only an \emph{approximation}
$V(t)$ to $U(t)$ which is good \cite{endnote22} for $|t| \le T$ then
the fourier transform of $g(t) = \frac{1}{2^n}\chi_T(t)\tr(V(t))$
will be close to that of $f(t) = \frac{1}{2^n}\chi_T(t)\tr(U(t))$.
One way to justify this is either to exploit the Parseval's
relation, i.e., that the fourier transform is a unitary operation on
$L_2$, or to use the result that if $\|f-g\|_1 \le \epsilon$ then
$\|\widehat{f} - \widehat{g}\|_{\infty} \le \epsilon$, where
$\|\cdot\|_1$ and $\|\cdot\|_{\infty}$ denote the $L_1$ and
$L_\infty$ norms, respectively.

We can also study the trace of the propagator in imaginary time:
consider $U(t+i\beta) = e^{-\beta H}e^{iHt}$. Taking the trace
yields
\begin{equation}
f(it+\beta) = \tr(U(t+i\beta)) = \sum_{j=0}^{2^n-1} e^{-\beta
E_j}e^{iE_jt}.
\end{equation}
We obtain $\mu_{H}$ from $f(it+\beta)$ for a fixed $\beta$ by
computing the laplace transform inversion integral
\begin{equation}
\mu_{H}(\omega) = \int_{\beta - i\infty}^{\beta + i\infty}ds\,
f(s)e^{s\omega}.
\end{equation}
For a fixed $\beta$ this can be done with an inverse fourier
transform in $t$: $\mu_{H}(\omega) =
e^{\beta\omega}\mathfrak{F}^{-1}[f(it+\beta)]$. It is
straightforward to see that if we only know an approximation
$V(t+i\beta)$ to $U(t+i\beta)$ valid for $|t| \le T$ then the
fourier transform $\widehat{g}(\omega)$ along $t$ of the cutoff
trace $g(it+\beta) = \chi_T(t)\tr(V(t+i\beta))$ provides a good
approximation to
$e^{-\beta\omega}(\widehat{X}_T\star\mu_{H})(\omega)$. Thus, after
normalisation, $e^{\beta\omega}\widehat{g}(\omega)$ provides a good
representation for the smeared eigenvalue density function for
$\omega\sim E_0$, where $E_0$ is the ground state energy. This
representation becomes exponentially worse as $\omega$ increases. By
combining this representation with the one obtained from pure time
evolution allows us to tradeoff the errors in the two
representations.

The preceding discussion serves to establish the fact that a good
approximation $V(t)$ to $U(t)=e^{iHt}$ which is valid for complex
times $|t|\le T$ provides sufficient information to resolve the
eigenvalue distribution function on a lengthscale $\delta \sim
O(\frac{1}{T})$. In the next part of this Letter we provide a
numerical algorithm, closely related to the DMRG, which efficiently
calculates a numerical representation for $(\widehat{\chi}_T\star
\mu_{H})(\omega)$.

The crucial idea underlying our numerical method is that a good
approximation $V(t)$ to the propagator $U(t)$ for a local $1$D
quantum spin lattice system can be stored efficiently (i.e.\ with
polynomial resources in $n$) on a classical computer for for $|t|
\le T$, where $T \sim O(\log(n))$ \cite{osborne:2005d} (see also
\cite{bravyi:2006a, eisert:2006b}). See Fig.~\ref{fig:qca} for an
illustration of the structure of the propagator $e^{itH}$ for an
arbitrary local spin system. This result allows us to \emph{certify}
that our algorithm can correctly obtain an approximate
representation for the smeared eigenvalue distribution function to
within a constant lengthscale which can be arbitrarily large (but
scaling at most logarithmically with $n$.)

The way we actually store a representation for $V(t)$ is as a
\emph{matrix product operator} (MPO) \cite{zwolak:2004a,
verstraete:2004b}. What we mean by this is that we represent an
operator $W\in \mathcal{B}(\mathcal{H})$ in the following fashion
\begin{equation}\label{eq:mpodef}
W = \sum_{\mathbf{j}\in Q_n}\mathbf{A}^{j_0}\mathbf{A}^{j_1}\cdots
\mathbf{A}^{j_{n-1}} \sigma^{j_0}\otimes \sigma^{j_1} \otimes
\cdots \otimes \sigma^{j_{n-1}},
\end{equation}
where $Q_n = (\mathbb{Z}/4\mathbb{Z})^{\times n}$, and
$\mathbf{A}^{j_0}$ and $\mathbf{A}^{j_{n-1}}$ are a collection of
four $C_0\times 1$ sized row vectors (respectively, four $1\times
D_{n-1}$ sized column vectors) and $\mathbf{A}^{j_k}$ are four
$C_k\times D_k$ sized matrices. Note that $C_{k+1} = D_{k}$. The
dimensions $C_k$ and $D_k$ are called the \emph{auxiliary
dimensions} for site $k$. It is clear that if the sizes of the
auxiliary dimensions are bounded by polynomials in $n$, i.e.\ $C_k
\le \text{poly}(n)$ and $D_k \le \text{poly}(n)$, then the operator
$W$ can be stored with polynomial resources in $n$. Also note that
all operators can be represented exactly as in Eq.~(\ref{eq:mpodef})
by taking the auxiliary dimensions to be large enough: $C_k = D_k =
2^n$ suffices.

\begin{figure}
\center
\includegraphics{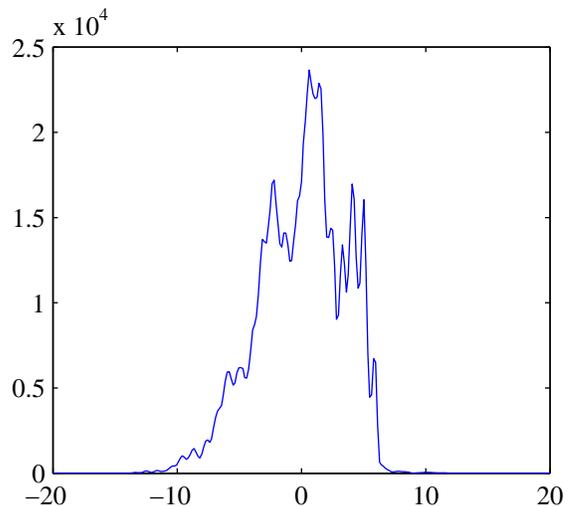}
\caption{The (scaled) eigenvalue density function for an instance of
the random antiferromagnetic Heisenberg model $H = \sum_{k=0}^{n-2}
J_k \boldsymbol{\sigma}_k\cdot \boldsymbol{\sigma}_{k+1}$, where the
couplings $J_k$ were chosen uniformly at random from the interval
$(0, 1)$, for a chain of $n=20$ spins. The eigenvalue density
function has been scaled so that the total area beneath the curve is
$2^{20}$, the total number of eigenvalues. The calculations were
performed with maximum auxiliary dimension $D=20$ and $T=20$. The
windowing gaussian has width $s=5$, thus the eigenvalue density
function is correct down to a lengthscale $2\pi/5$. The theoretical
worst-case error arising from the Trotter decomposition and
truncation of $D$ is $O(10^{-2})$ according to the $L_2$ and
$L_\infty$ norms.}\label{fig:gsm}
\end{figure}

\begin{figure}
\includegraphics{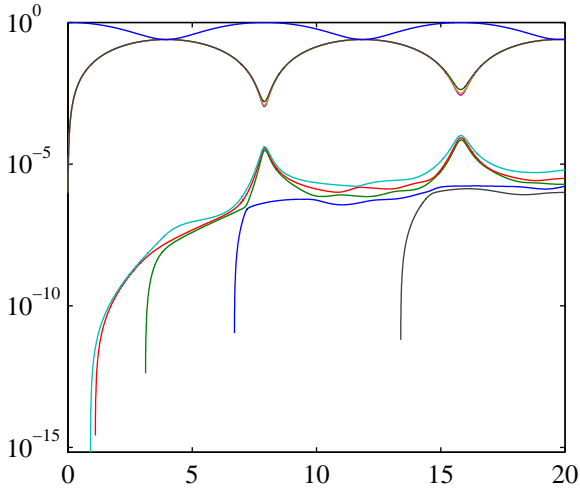}
\caption{The Schmidt coefficients as a function of time between the
first 10 spins and the second 10 spins for the MPO representation of
the propagator $e^{itH}$ for the random Heisenberg model $H$ studied
in Fig.~\ref{fig:gsm}. (The decay of the Schmidt coefficients
provide a measure of how difficult the propagator is to store
\cite{verstraete:2005a} and are calculated as the squares of the
singular values of $M_{\mathbf{j}\mathbf{k}} =
\mathbf{A}^{j_0}\mathbf{A}^{j_1}\cdots\mathbf{A}^{j_9}
\mathbf{A}^{k_0}\cdots\mathbf{A}^{k_{9}}$, where $V(t) =
\sum_{\mathbf{j}\in Q_{20}}\mathbf{A}^{j_0}\mathbf{A}^{j_1}\cdots
\mathbf{A}^{j_{19}} \sigma^{j_0}\otimes \sigma^{j_1} \otimes \cdots
\otimes \sigma^{j_{19}}$.) The Schmidt coefficients grow
approximately exponentially as a function of
time.}\label{fig:schmidt}
\end{figure}

We obtain the MPO representation for $V(t)$ via the following
method. First we break $H$ into two pieces $A$ and $B$ which contain
the interaction terms on the even (respectively, odd) sites. Note
that each term within $A$ (respectively $B$) commutes with all the
other terms within $A$ (respectively, $B$). Then we exploit the
Lie-Trotter expansion
\begin{equation}
e^{iHt} =
\lim_{m\rightarrow\infty}(e^{iA\frac{t}{m}}e^{iB\frac{t}{m}})^m,
\end{equation}
to write our expression for $V(t)$, i.e., we pick some $m$ and write
$V(t) = W^m$, where $W = e^{iA\frac{t}{m}}e^{iB\frac{t}{m}}$. We
next express $e^{ih_l\frac{t}{m}} = \sum_{\alpha, \beta =0}^3
c^{[j]}_{\alpha\beta} \sigma_j^{\alpha}\otimes
\sigma_{j+1}^{\beta}$. By writing $W$ in terms of the standard
product operator basis we obtain the MPO representation
\begin{equation}
W = \sum_{\mathbf{j}\in Q_n}\mathbf{B}^{j_0}\mathbf{A}^{j_1}\cdots
\mathbf{B}^{j_{n-1}} \sigma^{j_0}\otimes \sigma^{j_1} \otimes
\cdots \otimes \sigma^{j_{n-1}},
\end{equation}
where $B^{j_0}_\alpha = c_{j_0,\alpha}^{[0]}$, $\mathbf{A}^{j_k} =
\mathbf{g}^{j_k}\mathbf{c}^{[k]}$ on even sites, $\mathbf{A}^{j_k} =
\overline{\mathbf{g}}^{j_k}\mathbf{c}^{[k]}$ on odd sites, and
$B^{j_{n-1}}_\alpha = \delta_{j_{n-1},\alpha}$.

Now we show that if two MPO's $J$ and $K$ have maximum auxiliary
dimensions $D_J$ and $D_K$ then $JK$ is expressible as a MPO $L$
with auxiliary dimension $D_{JK} \le D_J D_K$. Representing $J =
\sum_{\mathbf{j}\in Q_n}\mathbf{A}^{j_0}\mathbf{A}^{j_1}\cdots
\mathbf{A}^{j_{n-1}} \sigma^{j_0}\otimes \sigma^{j_1} \otimes
\cdots \otimes \sigma^{j_{n-1}}$ and $K = \sum_{\mathbf{j}\in
Q_n}\mathbf{B}^{j_0}\mathbf{B}^{j_1}\cdots \mathbf{B}^{j_{n-1}}
\sigma^{j_0}\otimes \sigma^{j_1} \otimes \cdots \otimes
\sigma^{j_{n-1}}$ and taking the product gives $L = JK =
\sum_{\mathbf{j}\in Q_n}\mathbf{C}^{j_0}\mathbf{C}^{j_1}\cdots
\mathbf{C}^{j_{n-1}} \sigma^{j_0}\otimes \sigma^{j_1} \otimes
\cdots \otimes \sigma^{j_{n-1}}$, where
\begin{equation}
\mathbf{C}^{l_\gamma} = \sum_{j_\gamma,k_\gamma=0}^{3}
\mathbf{A}^{j_\gamma}\otimes \mathbf{B}^{k_\gamma} {g^{j_\gamma
k_\gamma}}_{l_\gamma}.
\end{equation}

We now apply this recipe to $W^m = V(t)$. Obviously, after a
couple of products, $W^m$ potentially requires an exponentially
large auxiliary dimension to represent it perfectly. It is here
that we use a method similar to the DMRG truncation to reduce the
size of this auxiliary dimension. We begin by representing $W^l$
as an MPO perfectly for as large an $l$ as possible. Then we
minimise the Hilbert-Schmidt norm difference $\| W^l - Y
\|_{\text{HS}} = \sqrt{\tr((W^l - Y)^\dag(W^l - Y))}$, where $Y$
is a MPO with a smaller auxiliary dimension. This is a
multiquadratic optimisation problem and can be solved numerically
in polynomial resources in $n$. (For a detailed description of
this procedure see \cite{verstraete:2004b}.) We then use the
approximation $Y$ to obtain an approximation $YW$ to $W^{l+1}$ and
repeat this process for the desired number of iterations.

Given an approximation $\widetilde{V}(t)=\sum_{\mathbf{j}\in
Q_n}\mathbf{A}^{j_0}(t)\mathbf{A}^{j_1}(t)\cdots
\mathbf{A}^{j_{n-1}}(t) \sigma^{j_0}\otimes \sigma^{j_1} \otimes
\cdots \otimes \sigma^{j_{n-1}}$ to $U(t)$ as an MPO with bounded
auxiliary dimension $D$ it is straightforward to obtain the trace
efficiently: $\tr(\sigma^{\alpha}) = \delta_{\alpha,0}$ gives
$\tr(\widetilde{V}(t)) = \mathbf{A}^{0}_0\mathbf{A}^{0}_1\cdots
\mathbf{A}^{0}_{n-1}$.

This procedure provides us with a discrete representation $g_k \sim
\tr(W^k)$, $k = 0, 1, \ldots, m-1$, for an approximation to the
trace $f(t_k)$ of $U(t_k)$. To obtain an approximate representation
for $\mu_{H}$ we apply a discrete fourier transform to
$\widetilde{g}_k = e^{-\frac{k^2}{2s^2}}g_k$ for some $s$ chosen
small enough to cutoff $g_k$ completely by $k_{\text{max}}$. A
standard result of fourier analysis shows that this discrete
representation will be a good representation for the smeared
eigenvalue density function as long as we sample more rapidly than
the Nyquist frequency $\nu$. We can provide a bound for $\nu$ by
noting that the largest eigenvalue $E_{\text{max}}$ of $H$ satisfies
$E_{\text{max}} \le \|H\|$, which is $O(n)$.

We have applied this numerical method to study the eigenvalue
distribution function for a random antiferromagnetic Heisenberg
model on $20$ spins, see Fig.~\ref{fig:gsm} and
Fig.~\ref{fig:schmidt}.

\begin{acknowledgements}
Helpful discussions with Jens Eisert and Frank Verstraete are
gratefully acknowledged.
\end{acknowledgements}


\begin{thebibliography}{20}
\bibitem{endnote21}{The eigenvalue density
function should not be confused with the \emph{local densiy of
states} (LDOS) which is the distribution $p_j = |c_j|^2$ of
coefficients of the state $A|E_0\rangle = \sum_{j=0}^{2^n-1}
c_j|E_j\rangle$ in the energy eigenbasis, where $A$ is an arbitrary
operator.}
\bibitem{endnote22}{By ``good'' we mean that, for some $\epsilon $,
$\delimiter "026B30D U(t)-V(t)\delimiter "026B30D  \le \epsilon $,
for $|t| \le T$, where $\delimiter "026B30D \cdot \delimiter
"026B30D $ denotes operator norm.}
\end{thebibliography}
\end{document}